\documentclass[12pt,preprint,iop,apj]{emulateapj}
\usepackage{graphicx}
\newcommand{\va}{v_{\mathrm{A}}}
\newcommand{\vai}{v_{\mathrm{Ai}}}
\newcommand{\vae}{v_{\mathrm{Ae}}}

\newcommand{\ta}{\tau_{\mathrm{A}}}

\newcommand{\pd}{\partial}

\newcommand{\vk}{v_{\rm k}}

\newcommand{\rhoi}{\rho_{\rm i}}
\newcommand{\rhoe}{\rho_{\rm e}}

\newcommand{\ld}{L_{\rm D}}

\begin{document}

	\title{RESONANTLY DAMPED PROPAGATING KINK WAVES IN LONGITUDINALLY STRATIFIED SOLAR WAVEGUIDES}

	\shorttitle{KINK WAVES IN STRATIFIED SOLAR WAVEGUIDES}

   \author{R. Soler$^1$, J. Terradas$^2$, G. Verth$^{1,3}$, and M. Goossens$^1$}
   \affil{$^1$Centre for Plasma Astrophysics, Department of Mathematics, Katholieke Universiteit Leuven,
              Celestijnenlaan 200B, 3001 Leuven, Belgium}
              \email{roberto.soler@wis.kuleuven.be}

 \affil{$^2$Departament de F\'isica, Universitat de les Illes Balears,
              E-07122, Palma de Mallorca, Spain}

\affil{$^3$School of Computing, Engineering and Information Sciences, Northumbria University, Newcastle Upon Tyne, NE1 8ST, England, UK}

  \begin{abstract}

It has been shown that resonant absorption is a robust physical mechanism to explain the observed damping of magnetohydrodynamic (MHD) kink waves in the solar atmosphere due to naturally occurring plasma inhomogeneity in the direction transverse to the direction of the magnetic field. Theoretical studies of this damping mechanism were greatly inspired by the first observations of post-flare \textit{standing} kink modes in coronal loops using the Transition Region And Coronal Explorer (TRACE). More recently, these studies have been extended to explain the attenuation of \textit{propagating} coronal kink waves observed by the Coronal Multi-Channel Polarimeter (CoMP). In the present study, for the first time we investigate the properties of propagating kink waves in solar waveguides including the effects of \textit{both} longitudinal and transverse plasma inhomogeneity. Importantly, it is found that the wavelength is only dependent on the longitudinal stratification and the amplitude is simply a product of the two effects. In light of these results the advancement of solar atmospheric magnetoseismology by exploiting high spatial/temporal resolution observations of propagating kink waves in magnetic waveguides to determine the length scales of the plasma inhomogeneity along and transverse to the direction of the magnetic field is discussed.

  \end{abstract}


%________________________________________________________________

\section{INTRODUCTION}

Resonant absorption, caused by plasma inhomogeneity in the direction transverse to
the magnetic field, has proved to be the most likely candidate for explaining the
observed attenuation of magnetohydrodynamic (MHD) kink waves in the solar atmosphere \citep[see][for a review about this damping mechanism]{goossens2006,goossensIAU}. E.g., resonant absorption is a feasible explanation for the damping of kink MHD waves in both coronal loops \citep{goossens2002} and in the fine structure of solar prominences \citep[see, e.g.,][and references therein]{oliver,arreguiballester}. Most of theoretical studies of
resonant absorption assume that the inhomogeneity is in the radial direction exclusively, and
only a few works \citep[e.g.,][]{andries2005, arregui2005,dymovaruderman2,andries2009a,solerfine}  have
studied the properties of resonantly damped standing kink waves when plasma
inhomogeneity is also present in the direction of the magnetic field.  These studies
have been restricted to standing waves, so the amplitude of
the standing wave is damped in time as a result of the resonance. Nevertheless, there are observational indications of
spatial damping of propagating waves along different waveguides in the
solar atmosphere. Some clear examples have been reported by
\citet{tomczyketal07} and \citet{tomczyk09} in coronal loops using Coronal Multi-Channel Polarimeter (CoMP) data. Other
observational evidence of propagating kink waves in wave guides of the solar
atmosphere are, e.g.,  \citet{depontetal07,he2009a,he2009b} in spicules, \citet{okamoto07}
 in the fine structure of prominences, and \citet{lin07,lin09} in filament threads. The theoretical modeling of the
spatial damping of traveling kink waves due to resonant absorption has been carried out
by \citet{TGV} and \citet{VTG} analytically, and by \citet{pascoeetal10,pascoeetal11} numerically. All these works
assumed a homogeneous density in the longitudinal direction. \citet{TGV} showed that the damping length by resonant absorption is inversely proportional to the wave frequency. It was predicted that high-frequency waves become damped in length scales smaller than low-frequency waves. \citet{VTG} showed that this theory is consistent with the CoMP observations reported by \citet{tomczyketal07} and \citet{tomczyk09}. This means that in solar waveguides resonant absorption acts as a natural low-pass filter.

Some recent investigations have extended the analytical work of \citet{TGV} by adding effects not considered in their original paper. \citet{solerspatial} took into account the effect of partial ionization in the single-fluid approximation, so kink waves are damped by both resonant absorption and ion-neutral collisions in their configuration. \citet{solerspatial} concluded that in thin tubes the effect of resonant absorption dominates, recovering the result of \citet{TGV} in the fully ionized case that the damping length is inversely proportional to the frequency. Therefore, the result of \citet{solerspatial} allows the theory developed by \citet{TGV} to be applied to kink waves in partially ionized plasmas as, e.g., chromospheric spicules and solar prominences.  Alternately, \citet{solerflow} have shown that in the presence of flow the damping length remains  inversely proportional to the frequency, but the factor of proportionality is different for forward and backward propagating waves to the flow direction and depends on the characteristics of the flow. However, none of these previous works investigated the effect of longitudinal stratification.

The analysis of kink MHD waves in longitudinally homogeneous models make some
problems more tractable from an analytical point of view. Realistic models of solar waveguides should include longitudinal stratification. E.g., coronal loops can extend out into the atmosphere up to heights of the order of several density scale heights. This means
that the differences in the density at their footpoints and apex can be
significant. There is evidence of such stratification in coronal loops both from emission measure analysis \citep[e.g.,][]{aschwanden1999} and coronal magnetoseismology \citep[see][for review of this topic]{andries2009b}. Regarding spicules, there have been spectroscopic line intensity studies to show the plasma density is longitudinally strongly stratified \citep[e.g.,][]{beckers1968, makita2003} and more recently by magnetoseismology \citep[][]{verth2011}. It is also clear from H$\alpha$ observations \citep[e.g.,][]{lin08} that prominence threads too exhibit longitudinal density inhomogeneity. For these
reasons, the main aim of the present paper is to investigate the effect of
longitudinal stratification on the properties of resonantly damped propagating kink
waves. This is an interesting problem from a theoretical, as well as
observational point of view. The reason is that a radial variation of density
causes damping due to resonant absorption while a longitudinal variation might cause
an increase of the amplitude. An increase of the amplitude due to stratification along the magnetic field direction has been obtained by, e.g., \citet{inekealfven} for phase mixed Alfv\'en waves and \citet{inekeslow} for propagating slow modes.

In this paper, we investigate the effect of longitudinal density variation on propagating kink waves by using both analytical and numerical approaches. Analytical theory is used to the largest possible extent. To do so, we apply standard approximations usually adopted in the preceding literature for the investigation of kink waves in magnetic tubes. The  approximations used in the analytical part of this paper are briefly discussed and justified in the next two paragraphs. In addition to the analytical theory, the problem is solved numerically beyond the limitations of the analytical approximations. The comparison between the analytical results and those from the full numerical simulations will enable us to check the validity of the analytical theory.

Our waveguide model is a straight cylindrical magnetic flux tube, inhomogeneous in both the radial and longitudinal directions, and embedded in a magnetized environment. To investigate propagating kink waves analytically we adopt the following approximations. We use the thin tube (TT) approximation, so that the wavelength, $\lambda$, is much larger than the radius of the tube, $R$. This condition is easily fulfilled for propagating waves observed in the solar atmosphere (see the various references given above). The variation of density in the radial direction is confined to a thin transitional layer, and we adopt the thin boundary (TB) approximation  \citep[see details in, e.g.,][]{hollwegyang,SGH91,goossens92,goossensIAU}. The TB approximation has been proved to be very accurate even when radial inhomogeneity is not restricted to a thin layer \citep[see, e.g.,][]{tom3, andries2005, arregui2005}. Regarding inhomogeneity in the longitudinal direction, we assume that the kink mode wavelength is shorter than the longitudinal inhomogeneity length scale, so that the WKB approximation can be applied.  In the numerical part of this paper, the problem is solved in the general case where large variations of density in both the radial and longitudinal directions are considered and for arbitrary wavelength. 	

Curvature is neglected in the model. The effect of curvature was studied by, e.g., \citet{tom1} and \citet{terradascurv}, and it has been reviewed by \citet{tom2}. Analytically, \citet{tom1} showed that curvature has no first-order effect on the frequency and the damping of kink modes in curved cylindrical models. The result that curvature has only a minor effect on kink MHD waves was also numerically confirmed by \citet{tom1} and \citet{terradascurv}. Wave damping due to leakage is not considered. Leaky waves, i.e., waves damped by MHD radiation, were first studied in magnetic flux tubes by \citet{spruit}, whose conclusion was that leakage is only important for wavelengths of the order of the tube radius or smaller. For thin tubes leakage is unimportant. This result was confirmed by \citet{goossenshollweg} and \citet{goossens2009}.  Finally, the $\beta = 0$ approximation is also adopted, where $\beta$ is the ratio of gas pressure to magnetic pressure. The assumption of $\beta = 0$ implies that waves do not have motions along the equilibrium magnetic field direction. It has been explained by, e.g., \citet{spruit} and \citet{goossens2009} that the longitudinal component of the velocity is proportional to the transverse component, with a factor of proportionality that depends of $\beta$ and $(R/\lambda)^2$. In the low-$\beta$ case and for thin tubes the longitudinal component of the velocity is much smaller than the transverse component. In magnetic structures of the solar atmosphere $\beta \ll 1$. This means that the longitudinal component of the velocity can be neglected and the $\beta = 0$ approximation can be safely adopted. For the effect of plasma $\beta$ on the resonant damping of kink waves, \citet{solerslow} \citep[see also][]{goossens2009} showed that the contribution of the slow continuum damping, present when $\beta \ne 0$, is negligible compared to the Alfv\'en continuum damping.

This paper is organized as follows. The description of the model configuration and basic equations are given in Section~\ref{sec:model}. Then, we obtain general expressions for the amplitude and wavelength of propagating kink waves in longitudinally and transversely inhomogeneous magnetic flux tubes in Section~\ref{sec:general}. Later, we use these general expressions to study resonantly damped propagating kink waves in stratified solar waveguides and compare the analytical predictions with full numerical time-dependent simulations in Section~\ref{sec:appli}. The implications for the method of solar atmospheric magnetoseismology are given in Section~\ref{sec:seis}. Finally, Section~\ref{sec:discussion} contains the discussion of the results and our conclusions.

\section{MODEL AND BASIC EQUATIONS}

\label{sec:model}

In the $\beta = 0$ approximation, the basic equations for the discussion of linear ideal MHD waves are,
\begin{eqnarray} \rho \frac{\pd {\bf v}}{\pd t} &=& \frac{1}{\mu} \left( \nabla
\times {\bf b} \right) \times {\bf B}, \label{eq:b1} \\ \frac{\pd {\bf b}}{\pd t} &=&
\nabla \times \left( {\bf v} \times {\bf B} \right), \label{eq:b2} \end{eqnarray}
where $\rho$ is the plasma density, ${\bf B}$ is the equilibrium magnetic field, ${\bf v}$ is the velocity perturbation, ${\bf b}$ is the magnetic field perturbation, and $\mu$ is the magnetic permittivity. We use linear theory in our investigation because the Alfv\'{e}n velocity in the solar corona is of the order of 1000~km~s$^{-1}$, while the peak-to-peak velocity amplitude of the propagating waves observed using CoMP is only about 1~km~s$^{-1}$ \citep[see][]{tomczyketal07, tomczyk09}.  We study MHD waves in a static background, so the equilibrium flow is absent from our discussion. The effect of flow on resonantly damped propagating kink waves has been recently studied by \citet{solerflow}.

The equilibrium configuration is a straight cylindrical magnetic flux tube of
average radius $R$ embedded in a magnetized plasma environment. For convenience, we use
cylindrical coordinates, namely $r$, $\varphi$, and $z$ for the radial, azimuthal,
and longitudinal coordinates, respectively. The axis of the cylinder is set along the
$z$-direction. In the following expressions, subscripts $\rm i$ and $\rm e$ refer
to the internal and external plasmas, respectively.  We denote by $\rhoi (z)$ and
$\rhoe (z)$ the internal and external densities, respectively. Both of these quantities are
functions of $z$. There is a nonuniform transitional layer in the transverse
direction that continuously connects the internal density to the external density.
The thickness of the layer is $l$, and extends in the interval $R - l/2 \leq r \leq R
+ l/2$. The equilibrium magnetic field is straight and homogeneous, ${\bf B} = B
\hat{e}_z$, with $B$ constant. As $\varphi$ is an ignorable coordinate in our model
and we consider waves propagating along the tube with a fixed frequency, we write all
perturbations proportional to $\exp \left( i m \varphi - i \omega t \right)$, where
$m$ is the azimuthal wavenumber and $\omega$ is the wave (angular) frequency. The
frequency $f$ is related to $\omega$ by $\omega = 2\pi f$. Kink modes,
i.e., the only wave modes that can displace the magnetic cylinder axis and so
produce transverse motions of the whole flux tube \citep[see,
e.g.,][]{edwinroberts,goossens2009}, are described by $m=1$. Due to the presence of
a transverse inhomogeneous transitional layer, wave modes with $m \neq 0$ are
spatially damped by resonant absorption. In the position within the transverse
inhomogeneous transitional layer where the wave frequency coincides with the local
Alfv\'en frequency, i.e., the resonance position, the wave character changes and
becomes more Alfv\'enic as the wave propagates. As a result of this process,
transverse motions of the flux tube are damped while azimuthal motions within
the transitional layer are amplified.

 We use the TT approximation, i.e., $\lambda \gg R$, where $\lambda$ is the wavelength along the tube, and apply the formalism of \citet{dymovaruderman,dymovaruderman2}.  The effect of the resonant damping is included here by introducing jump conditions for perturbations at the resonance position along with the TB approximation \citep[see details in, e.g.,][]{hollwegyang,SGH91,goossens92}. The combination of both TT and TB approximations is usually called the TTTB approach \citep[see the recent review by][]{goossensIAU}. Then, following the scaling arguments of \citet{dymovaruderman,dymovaruderman2}, the total pressure perturbation $P = B b_z / \mu$, with $b_z$ the longitudinal component of the magnetic field perturbation, evaluated at both sides of the transverse transitional layer can be cast as
\begin{eqnarray}
P_{\rm i} &\approx& \left( \frac{r}{R} \right)^m A_{\rm i} (z), \quad \textrm{at} \quad r \approx R - \frac{l}{2}, \label{eq:pi} \\
P_{\rm e} &\approx& \left( \frac{R}{r} \right)^m A_{\rm e} (z), \quad \textrm{at} \quad r \approx R +\frac{l}{2}, \label{eq:pe}
\end{eqnarray}
with $A_{\rm i} (z)$ and $A_{\rm e} (z)$ arbitrary functions of $z$. In the TB approximation, the condition of $P$ continuity across the resonant layer \citep[e.g.,][]{SGH91,goossens92,goossens95} implies $P_{\rm i} = P_{\rm e}$ at $r = R$, which means that $A_{\rm i} (z) = A_{\rm e} (z) = A (z)$.

On the other hand, from Equations~(\ref{eq:b1}) and (\ref{eq:b2}) we obtain the following relation
\begin{equation}
  \frac{\pd^2 v_r}{\pd t^2} - \va^2 (z) \frac{\pd^2 v_r}{\pd z^2} = - \frac{1}{\rho (z)} \frac{\pd^2 P}{\pd r \pd t}, \label{eq:vrp}
\end{equation}
where $v_r$ is the radial velocity perturbation, and $\va^2 (z) = \frac{B^2}{\mu \rho (z)}$ is the Alfv\'en velocity squared. After using Equations~(\ref{eq:pi}) and (\ref{eq:pe}) in Equation~(\ref{eq:vrp}) we arrive at the two following expressions,
\begin{eqnarray}
 \vai^2(z) +  \frac{\pd^2 v_{r \rm i}}{\pd z^2} + \omega^2 v_{r \rm i} =- i \frac{\omega}{\rhoi (z)} \frac{m}{R} A (z), \quad &\textrm{at}& \quad r \approx R - \frac{l}{2}, \label{eq:vri} \\
\vae^2(z) +  \frac{\pd^2 v_{r \rm e}}{\pd z^2} + \omega^2 v_{r \rm e} =  i \frac{\omega}{\rhoe (z)} \frac{m}{R} A (z), \quad &\textrm{at}& \quad r \approx R + \frac{l}{2},\label{eq:vre}
\end{eqnarray}
where $v_{r \rm i}$ and $v_{r \rm e}$ are the radial velocity perturbations at $r \approx R - l/2$ and $r \approx R + l/2$, respectively. Now, we take into account that in the TB approximation the jump of $v_r$ across the resonance layer, namely $[v_r]$, is assumed to be the same as the jump of $v_r$ across the whole transverse inhomogeneous layer, i.e., $v_{r \rm e} - v_{r \rm i} = [v_r]$. For a straight and constant magnetic field, $[v_r]$ is given by \citep[see, e.g.,][]{SGH91,goossens92,goossens95,tirry,andries2005,dymovaruderman2},
\begin{equation}
 [v_r] = -\pi \frac{m^2 / R^2}{\omega \left| \pd_r \rho \right|} P, \qquad \textrm{at} \qquad r \approx R, \label{eq:jump}
\end{equation}
where $\left| \pd_r \rho \right|$ is the radial derivative of the transverse density profile at the resonance position, which is assumed to take place at $r \approx R$. This is a reasonable assumption in the TTTB approximation. Subsequently, we combine Equations~(\ref{eq:vri}) and (\ref{eq:vre}) to eliminate function $A(z)$ and use the jump condition (Equation~(\ref{eq:jump})) to arrive at a single equation for $v_{r \rm i}$ and $[v_r]$. Forthwith, we denote $v_{r \rm i}$ as $v_r$ to simplify the notation and the resultant expression is
\begin{equation}
 \frac{\pd^2 v_r}{\pd z^2} + \frac{\omega^2}{\vk^2 (z)} v_r = - \frac{1}{2} \left( \frac{\pd^2 [v_r]}{\pd z^2} + \frac{\omega^2}{\vae^2 (z)} [v_r]  \right),\label{eq:basic}
\end{equation}
where
\begin{equation}
 \vk^2 (z) = \frac{2 B^2}{\mu \left( \rhoi (z) + \rhoe (z) \right)}, \label{eq:vk}
\end{equation}
is the squared of the kink velocity, i.e, the phase velocity of the propagating kink wave. For future use, it is instructive to note that the kink velocity $\vk$ is a function of $z$. Finally, we use Equation~(\ref{eq:vri}) along with the result that $P$ is constant across the resonant layer to relate $[v_r]$ with $v_r$, obtaining
\begin{equation}
  [v_r] \approx - i \pi \frac{m}{R} \frac{\rhoi (z)}{\left| \pd_r \rho \right|} \left( \frac{\vai^2 (z)}{\omega^2}\frac{\pd^2 v_r}{\pd z^2} + v_r \right), \label{eq:press}
\end{equation}
which we put in Equation~(\ref{eq:basic}) to get an equation involving $v_r$ only, namely
\begin{equation}
 C_4 \frac{\partial^4 v_r}{\partial z^4} + C_3 \frac{\partial^3 v_r}{\partial z^3} + C_2 \frac{\partial^2 v_r}{\partial z^2} + C_1 \frac{\partial v_r}{\partial z} + C_0 v_r = 0, \label{eq:vrfull}
\end{equation}
with
\begin{eqnarray}
 C_4 &=& - i \frac{\pi}{2} \frac{m}{R} \frac{1}{\omega^2} \frac{\rhoi (z)}{\left| \partial_r \rho \right|} \vai^2 (z), \nonumber \\
C_3 &=& - i \pi  \frac{m}{R} \frac{\rhoi (z)}{\omega^2} \frac{\partial}{\partial z} \left( \frac{1}{\left| \partial_r \rho \right|} \right) \vai^2 (z), \nonumber \\
C_2 &=& 1 - i  \frac{\pi}{2}\frac{m}{R} \left[ \frac{\rhoi (z)}{\omega^2}  \frac{\partial^2}{\partial z^2} \left( \frac{1}{\left| \partial_r \rho \right|} \right) \vai^2 (z) + \frac{\rhoi(z) + \rhoe(z)}{\left| \partial_r \rho \right|}  \right],\nonumber \\
C_1 &=& - i \pi \frac{m}{R}   \frac{\partial}{\partial z} \left( \frac{\rhoi(z)}{\left| \partial_r \rho \right|} \right), \nonumber \\
C_0 &=& \frac{\omega^2}{\vk^2(z)} - i  \frac{\pi}{2}\frac{m}{R} \left[ \frac{\partial^2}{\partial z^2} \left( \frac{\rhoi(z)}{\left| \partial_r \rho \right|} \right) + \frac{\omega^2}{\vae^2 (z)} \frac{\rhoi(z)}{\left| \partial_r \rho \right|}  \right] \label{eq:c2}.
\end{eqnarray}
Equation~(\ref{eq:vrfull}) is the main equation of this investigation. Note that the imaginary parts of the coefficient functions $C_0$--$C_4$ are all proportional to $m$.

\subsection{Case Without Longitudinal Stratification}

As a limiting case of Equation~(\ref{eq:vrfull}), we study wave propagation in a magnetic tube which is uniform in the longitudinal direction and try to recover the results of \citet[hereafter TGV]{TGV}. In order to do so, we take $\rhoi$ and $\rhoe$ constants and neglect their derivatives in the longitudinal direction. Since the background is independent of $z$ we can Fourier-analyze in the $z$-direction and write $v_r$ proportional to $\exp \left( i k_z z \right)$, with $k_z$ the longitudinal wavenumber. Equation~(\ref{eq:vrfull}) becomes
\begin{equation}
 C_4 k_z^4 - C_2 k_z^2 + C_0 = 0.
\end{equation}
Since we are investigating spatial damping, we write $k_z = k_{z \rm R} + i k_{z \rm I}$, approximate $k_{z \rm R}^2 \approx \omega^2/\vk^2$, and assume $k_{z \rm I} \ll k_{z \rm R}$ (weak damping), so we neglect terms with $k_{z \rm I}^2$. Finally the following relation is obtained,
\begin{equation}
 \frac{k_{z \rm I}}{k_{z \rm R}} = \frac{\pi}{8} \frac{m}{R} \frac{1}{\left| \partial_r \rho \right|} \frac{\left( \rhoi - \rhoe \right)^2}{\rhoi + \rhoe}, \label{eq:relkzrkzi}
\end{equation}
which coincides with Equation~(10) of TGV. Therefore, we recover the results of TGV for propagation and damping of kink waves in longitudinally homogeneous tubes. The reader is referred to TGV for a complete analysis of this case. It is worth noting that Equation~(\ref{eq:relkzrkzi}) also agrees with Equation~(20) of \citet{solerspatial}, obtained for the spatial damping of propagating kink waves in partially ionized threads of solar prominences, if the effect of damping by ion-neutral collisions is removed from their expression.

\section{GENERAL FORMULATION}

\label{sec:general}

\subsection{No Resonant Damping}

In the present and next Sections, we investigate the effects of longitudinal density stratification. First, we study the case when resonant damping is absent to assess the effect of density stratification on the amplitude of the propagating kink wave. Damping will be considered later. Then, Equation~(\ref{eq:vrfull}) reduces to
\begin{equation}
 \frac{\pd^2 v_r}{\pd z^2} + \frac{\omega^2}{\vk^2 (z)} v_r = 0. \label{eq:basicnores}
\end{equation}
Note that Equation~(\ref{eq:basicnores}) is independent of the azimuthal wavenumber, $m$. In the absence of damping, the waves are degenerate with respect to $m$ as long as $m \ne 0$.

To find a solution to Equation~(\ref{eq:basicnores}), we use the Wentzel-Kramers-Brillouin (WKB) approximation \citep[see, e.g.,][for details about the method]{bender}. We define the dimensionless longitudinal coordinate as,
\begin{equation}
 \zeta = \frac{z}{\Lambda}, \label{eq:zeta}
\end{equation}
where $\Lambda$ is the longitudinal inhomogeneity scale height, and express the solution of Equation~(\ref{eq:basicnores}) in the following form,
\begin{equation}
 v_r (\zeta) \approx Q \left( \zeta \right) \exp \left[ i \Lambda  \mathcal{K} \left( \zeta \right) \right], \label{eq:wkb}
\end{equation}
with $Q \left( \zeta \right)$ and $\mathcal{K} \left( \zeta \right)$ functions to be determined. The validity of the WKB approximation, and so of Equation~(\ref{eq:wkb}), is restricted to large values of  $\Lambda$, i.e., weak inhomogeneity, such as $\lambda/ \Lambda \ll 1$. We combine Equations~(\ref{eq:basicnores}) and (\ref{eq:wkb}), and separate the different terms according to their order with respect to $\Lambda$. As $\Lambda$ is large, the dominant terms are those with the highest order in $\Lambda$. From the leading term, i.e., that with $\mathcal{O} \left( \Lambda^2 \right)$, we obtain the following expression,
\begin{equation}
 \tilde{k}_z^2 (\zeta) = \frac{\omega^2}{\vk^2(\zeta)}, \label{eq:realkz}
\end{equation}
where $\tilde{k}_z (\zeta)$ has been defined as
\begin{equation}
 \tilde{k}_z (\zeta) \equiv \frac{{\rm d} \mathcal{K} (\zeta)}{{\rm d} \zeta}. \label{eq:defkz}
\end{equation}
We see that $\tilde{k}_z (\zeta)$ plays the role of the longitudinal wavenumber, which in our case is a function of $\zeta$ through the dependence of the kink velocity on $\zeta$ (Equation~(\ref{eq:vk})). Therefore, the wavelength, $\lambda (\zeta)$, is
\begin{equation}
\lambda (\zeta) = \frac{2\pi}{\tilde{k}_{z\rm R} (\zeta)} = \frac{2\pi}{\omega} \vk (\zeta) = \tau \, \vk (\zeta), \label{eq:wavelength}
\end{equation}
with $\tau$ the wave period. We obtain that $\lambda (\zeta)$ changes in $\zeta$ due to stratification. Therefore, function $\mathcal{K} (\zeta)$ is
\begin{equation}
 \mathcal{K} (\zeta) = \omega \int_0^\zeta \frac{1}{\vk (\xi)} {\rm d} \xi,
\end{equation}
where we have assumed $ \mathcal{K} (0) = 0 $.

On the other hand, from the contribution of  $\mathcal{O} \left( \Lambda^1 \right)$ in the WKB expansion of  Equation~(\ref{eq:basicnores}), we get the following expression
\begin{equation}
 \frac{\pd Q (\zeta) }{\pd \zeta} + \frac{1}{2 \tilde{k}_{z} (\zeta)} \frac{ \pd \tilde{k}_{z} (\zeta) }{\pd \zeta} Q (\zeta) = 0, \label{eq:ampli}
\end{equation}
which can be easily integrated to
\begin{equation}
 Q (\zeta) = Q (0) \left( \frac{\vk (\zeta)}{\vk (0)} \right)^{1/2}. \label{eq:amp}
\end{equation}
Equation~(\ref{eq:amp}) tells us how the amplitude changes with height due to density stratification. Density stratification modifies both the amplitude and the wavelength of the kink wave as it propagates along the magnetic tube. If density decreases with height, and hence the kink velocity increases with height, Equations~(\ref{eq:amp}) and (\ref{eq:wavelength}) show that both the wavelength and the amplitude increase.

 The complete expression for $v_r$ in the WKB approximation using the original variable $z$ is
\begin{equation}
 v_r (z) \approx v_r (0)  \left( \frac{\vk (z)}{\vk (0)} \right)^{1/2} \exp \left( i \omega \int_0^z \frac{1}{\vk (\tilde{z})} {\rm d} \tilde{z} \right) . \label{eq:vrassym}
\end{equation}

\subsection{Effect of Resonant Damping}

Hereafter, we include the effect of damping by resonant absorption and we try to solve the full Equation~(\ref{eq:vrfull}). As a completely general solution to Equation~(\ref{eq:vrfull}) is very difficult to obtain, we restrict ourselves to the case in which the internal and external densities are proportional. We define the density contrast, $\chi$, as
\begin{equation}
 \chi = \frac{\rhoi (z)}{\rhoe (z)}.\label{eq:contrast}
\end{equation}
We assume that parameter  $\chi$ is a constant independent of $z$. In addition, we write $\left| \partial_r \rho \right|$ as
\begin{equation}
 \left| \partial_r \rho \right| = \mathcal{F} \frac{\pi^2}{4} \frac{\rhoi(z) - \rhoe(z)}{l} =  \mathcal{F} \frac{\pi^2}{4} \frac{\chi - 1}{\chi} \frac{\rhoi(z)}{l}, \label{eq:deriv}
\end{equation}
with $\mathcal{F}$ a factor that depends on the form of the transverse density profile. For example, $\mathcal{F} = 4/\pi^2$ for a linear profile \citep{goossens2002} and $\mathcal{F} = 2/\pi$ for a sinusoidal profile \citep{rudermanroberts}. Then, coefficients $C_0$--$C_4$ in Equations~(\ref{eq:c2}) simplify to
\begin{eqnarray}
 C_4 &=& - i  \frac{2}{\pi}  \frac{l}{R} \frac{m}{\mathcal{F}} \frac{\chi}{\chi-1} \frac{1}{\omega^2}   \vai^2 (z), \nonumber \\
C_3 &=& - i  \frac{4}{\pi}  \frac{l}{R}\frac{m}{\mathcal{F}} \frac{\chi}{\chi-1} \frac{1}{\omega^2}  \frac{\pd \vai^2 (z)}{\pd z }, \nonumber \\
C_2 &=& 1 + i   \frac{2}{\pi} \frac{l}{R}  \frac{m}{\mathcal{F}} \left[ \frac{\chi}{\chi-1} \frac{1}{\omega^2} \frac{\pd^2 \vai^2 (z)}{\pd z^2 }  - \frac{\chi + 1}{\chi -1}  \right],\nonumber \\
C_1 &=& 0, \nonumber \\
C_0 &=& \frac{\omega^2}{\vk^2(z)} - i \frac{2}{\pi}  \frac{l}{R}   \frac{m}{\mathcal{F}} \frac{1}{\chi-1} \frac{\omega^2}{\vai^2 (z)}. \label{eq:c22}
\end{eqnarray}

As before, we use the WKB approximation to solve Equation~(\ref{eq:vrfull}) with the coefficients given in Equations~(\ref{eq:c22}). Then, we re-introduce the dimensionless longitudinal coordinate $\zeta = z/\Lambda$ (Equation~(\ref{eq:zeta})) and write $v_r (\zeta)$ in the form given in Equation~(\ref{eq:wkb}). Following the same process as in the case without resonant damping, we separate the different terms of Equation~(\ref{eq:vrfull})  according to their order with respect to $\Lambda$. From the leading contribution, i.e., that with $\mathcal{O} \left( \Lambda^4 \right)$, we obtain the following expression,
\begin{eqnarray}
 &&\tilde{k}_z^2 (\zeta) - \frac{\omega^2}{\vk^2(\zeta)} +  i \frac{\pi}{2} \frac{m}{R} \frac{\mu}{B^2} \frac{\rhoi (\zeta) \rhoe (\zeta)}{\left| \partial_r \rho \right|} \nonumber \\
&\times& \frac{\left(\omega^2 - \tilde{k}_z^2 (\zeta) \vai^2(\zeta) \right) \left(\omega^2 - \tilde{k}_z^2 (\zeta) \vae^2(\zeta) \right)}{\omega^2} = 0. \label{eq:ol4}
\end{eqnarray}
where $\tilde{k}_z (\zeta)$ is defined in Equation~(\ref{eq:defkz}). For convenience, we have used Equation~(\ref{eq:deriv}) to introduce $\left| \partial_r \rho \right|$ again, in order to illustrate that Equation~(\ref{eq:ol4}) is formally equivalent to Equation~(6) of TGV if the longitudinal dependence of the density is added to their expression. We follow the analytical process described in TGV and write $ \tilde{k}_z (\zeta)= \tilde{k}_{z\rm R} (\zeta)+ i \tilde{k}_{z\rm I} (\zeta)$. The effect of resonant absorption is to introduce an imaginary part of $\tilde{k}_z (\zeta)$, while we can reasonably assume that $\tilde{k}_{z\rm R} (\zeta)$ is the same as in the case without resonant damping, i.e., $ \tilde{k}_{z\rm R} (\zeta) \approx \omega/\vk (\zeta)$ according to Equation~(\ref{eq:realkz}).  Next, we assume weak damping, i.e., $\tilde{k}_{z\rm I} (\zeta) \ll \tilde{k}_{z\rm R} (\zeta)$, so we neglect terms with $\tilde{k}_{z\rm I}^2 (\zeta)$ in Equation~(\ref{eq:ol4}). Finally, a relation between $\tilde{k}_{z\rm R} (\zeta)$ and $\tilde{k}_{z\rm I} (\zeta)$ is derived, namely
\begin{equation}
 \frac{\tilde{k}_{z\rm I} (\zeta)}{\tilde{k}_{z\rm R} (\zeta)} = \frac{\pi}{8} \frac{m}{R} \frac{1}{\left| \partial_r \rho \right|} \frac{\left( \rhoi (\zeta) - \rhoe (\zeta)\right)^2}{\rhoi(\zeta)+ \rhoe(\zeta)}, \label{eq:relkzrkzi2}
\end{equation}
which coincides with our Equation~(\ref{eq:relkzrkzi}) and with Equation~(10) of TGV if the dependence of the densities in the longitudinal direction is explicitly included. Taking into account the definitions of Equations~(\ref{eq:contrast}) and (\ref{eq:deriv}), we rewrite Equation~(\ref{eq:relkzrkzi2}) as
\begin{equation}
 \frac{\tilde{k}_{z\rm I} (\zeta)}{\tilde{k}_{z\rm R} (\zeta)} = \frac{1}{2\pi} \frac{m}{\mathcal{F}}  \frac{l}{R} \frac{ \chi - 1}{\chi + 1}, \label{eq:relkzrkzi3}
\end{equation}
and we clearly see that the right-hand side of Equation~(\ref{eq:relkzrkzi3}) is independent of $\zeta$. We use Equation~(\ref{eq:wavelength}) and define the damping length by resonant absorption as $\ld (\zeta) = 1/\tilde{k}_{z\rm I} (\zeta)$, so we can obtain from Equation~(\ref{eq:relkzrkzi3}) that
\begin{equation}
 \frac{\ld (\zeta)}{\lambda (\zeta)} =  \frac{\mathcal{F}}{m} \frac{R}{l} \frac{\chi + 1}{ \chi - 1 }. \label{eq:dampratio}
\end{equation}
Again, note that the right-hand side of Equation~(\ref{eq:dampratio}) is independent of $\zeta$. Finally, we consider the expression for $\lambda (\zeta)$ of Equation~(\ref{eq:wavelength}) to give the expression for the damping length as
\begin{equation}
 \ld (\zeta) = 2 \pi \frac{\mathcal{F}}{m} \frac{R}{l} \frac{\chi + 1}{ \chi - 1 }\frac{1}{\omega} \vk (\zeta) \label{eq:damplen},
\end{equation}
which is equivalent to Equation~(22) of TGV. The damping length is inversely proportional to both $\omega$ and $m$. By using Equations~(\ref{eq:wavelength}) and (\ref{eq:damplen}), we compute the real and imaginary parts of $\mathcal{K}(\zeta) = \mathcal{K}_{\rm R}(\zeta) + i \mathcal{K}_{\rm I}(\zeta)$, namely
\begin{eqnarray}
 \mathcal{K}_{\rm R} (\zeta) &=& \int_0^\zeta \frac{2 \pi}{\lambda \left( \xi \right)} {\rm d}\xi =  \omega \int_0^\zeta \frac{1}{\vk (\xi)} {\rm d} \xi, \label{eq:mkre} \\
\mathcal{K}_{\rm I} (\zeta) &=& \int_0^\zeta \frac{1}{ \ld \left( \xi \right)} {\rm d}\xi \nonumber \\
 &=& \frac{1}{2\pi} \frac{m}{\mathcal{F}} \frac{l}{R} \frac{\chi- 1}{ \chi + 1 } \omega \int_0^\zeta \frac{1}{\vk (\xi)} {\rm d} \xi, \label{eq:mkim}
\end{eqnarray}
where again we assumed $\mathcal{K}_{\rm R} (0) =\mathcal{K}_{\rm I} (0) = 0$.

On the other hand, we get an equation for $Q (\zeta)$ from the contribution of  $\mathcal{O} \left( \Lambda^3 \right)$ in the WKB expansion of  Equation~(\ref{eq:vrfull}), namely
\begin{equation}
 \frac{\pd Q (\zeta) }{\pd \zeta} + \frac{1}{2 \tilde{k}_{z\rm R} (\zeta)} \frac{ \pd \tilde{k}_{z\rm R} (\zeta) }{\pd \zeta} Q (\zeta) = 0, \label{eq:ampli2}
\end{equation}
which  coincides with Equation~(\ref{eq:ampli}). Therefore, the solution to Equation~(\ref{eq:ampli2}) is given by Equation~(\ref{eq:amp}), meaning that the change of the amplitude in $\zeta$ due to density stratification is the same as in the case without resonant damping.

Finally, using Equations~(\ref{eq:amp}), (\ref{eq:mkre}) and (\ref{eq:mkim}) in Equation~(\ref{eq:wkb}), and returning to the original variable $z$, we obtain
\begin{equation}
 v_r (z) \approx \mathcal{A} (z) \exp \left( i \omega \int_0^z \frac{1}{\vk (\tilde{z})} {\rm d} \tilde{z} \right), \label{eq:vrexp}
\end{equation}
with $\mathcal{A} (z)$ the amplitude as a function of $z$ given by
\begin{eqnarray}
 \mathcal{A} (z) &=& v_r\left( 0 \right)  \left( \frac{\vk (z)}{\vk (0)} \right)^{1/2}   \nonumber \\
&\times& \exp \left( - \frac{1}{2\pi} \frac{m}{\mathcal{F}} \frac{l}{R} \frac{\chi- 1}{ \chi + 1 } \omega \int_0^z \frac{1}{\vk (\tilde{z})} {\rm d} \tilde{z} \right). \label{eq:amptot}
\end{eqnarray}
Equation~(\ref{eq:amptot}) shows that there are two effects that determine the amplitude of the propagating kink wave, i.e, longitudinal density stratification and  resonant absorption. While the effect of longitudinal stratification depends on whether the density increases or decrease with $z$, the role of resonant absorption is always to damp the wave and so to decrease its amplitude. It is clear from Equation~(\ref{eq:amptot}) that resonant absorption acts as a natural filter for waves with high frequencies and large $m$.

\section{APPLICATION TO PROPAGATING KINK WAVES IN STRATIFIED CORONAL LOOPS}

\label{sec:appli}

\subsection{Analytical Expressions}

Here, we apply the general theory described in the last Section to the case of driven kink waves in a stratified flux tube. For a particular application we model a coronal loop being driven at one footpoint. Assuming the longitudinal stratification of the coronal loop is symmetric about its apex, we focus our analysis on the loop half that contains the footpoint driver. We take the plasma to be more dense at the loop's footpoints to mimic a gravitationally stratified solar atmosphere and assume an exponential $z$ dependence in density, namely
\begin{eqnarray}
 \rhoi (z) &=& \rho_{\rm i 0} \exp \left( - \frac{z}{\Lambda} \right), \label{eq:densi} \\
\rhoe (z) &=& \rho_{\rm e 0} \exp \left( - \frac{z}{\Lambda} \right),\label{eq:dense}
\end{eqnarray}
where $\rho_{\rm i 0}$ and $\rho_{\rm e 0}$ are the internal and external
densities, respectively, at the footpoints of loop, i.e., at $z=0$, and $\Lambda$ is the scale height defined as
\begin{equation}
 \Lambda = \frac{L / 2}{\ln \left( \rho_{\rm i  0} /  \rho_{\rm apex} \right)}, \label{eq:deflamstrat}
\end{equation}
with $L$ the total length of the loop and $\rho_{\rm apex}$ the internal density at the apex, i.e., at $z = L/2$. As we assume the same stratification within the tube and in the corona, $\chi = \rhoi (z) / \rhoe(z) = \rho_{\rm i 0} / \rho_{\rm e 0}$ is a constant. According to Equations~(\ref{eq:densi}) and (\ref{eq:dense}), the square of the kink velocity is
\begin{equation}
 \vk^2 (z) = v_{\rm k 0}^2 \exp \left(  \frac{z}{\Lambda} \right), \label{eq:kinkspeed}
\end{equation}
where $v_{\rm k 0}^2 = 2 B^2/\left[\mu \left( \rho_{\rm i  0} +   \rho_{\rm e  0} \right)\right]$ is the square of the kink velocity at $z=0$. As in our model the magnetic field is straight and constant, the density scale height and the kink velocity scale height are proportional. CoMP observations of coronal propagating waves \citep{tomczyk09} show that the kink velocity is relatively constant with height so that the kink velocity scale height is much longer that the typical coronal scale height. The effect of flux tube expansion might be responsible for keeping the kink velocity relatively constant.  The effect of flux tube expansion is discussed in Section~\ref{sec:seis}.

\subsubsection{Case Without Resonant Damping}

First, we analyze the case without resonant damping. The kink mode propagation is governed by Equation~(\ref{eq:basicnores}). The expression for $\vk^2 (z)$ given in Equation~(\ref{eq:kinkspeed}) enable us to rewrite Equation~(\ref{eq:basicnores}) as a Bessel Equation of order 0. Then, it is possible to obtain an exact solution of Equation~(\ref{eq:basicnores}) as
\begin{eqnarray}
  v_r (z) &=& A_1 H_0^{\rm (1)} \left[ 2 \Lambda \frac{\omega}{v_{\rm k 0}} \exp \left(- \frac{z}{2 \Lambda} \right) \right] \nonumber \\
 &+& A_2 H_0^{\rm (2)} \left[ 2 \Lambda \frac{\omega}{v_{\rm k 0}} \exp \left(- \frac{z}{2 \Lambda} \right)  \right], \label{eq:vrgen}
\end{eqnarray}
with $H_0^{\rm (1)}$ and $H_0^{\rm (2)}$ the Hankel functions of the first and second kind of order 0, and $A_1$ and $A_2$ complex constants. The real part of Equation~(\ref{eq:vrgen}) contains the solution with physical meaning. The functions $H_0^{\rm (1)}$ and $H_0^{\rm (2)}$ represent waves propagating towards the positive and negative $z$-directions, respectively.  By setting $A_2 = 0$, we restrict ourselves to propagation towards the positive $z$-direction. In addition, we chose $A_1$ so that $\Re \left( v_r \right) = 1$ and $\Re \left( \frac{\partial v_r}{\partial z} \right) = 0$ at $z=0$, hence
\begin{equation}
 A_1  = \frac{Y_1 \left( 2 \Lambda \frac{\omega}{{\vk}_0} \right) + i J_1 \left( 2 \Lambda \frac{\omega}{{\vk}_0} \right)}{J_0 \left( 2 \Lambda \frac{\omega}{{\vk}_0} \right) Y_1 \left( 2 \Lambda \frac{\omega}{{\vk}_0} \right) - J_1 \left( 2 \Lambda \frac{\omega}{{\vk}_0} \right) Y_0 \left( 2 \Lambda \frac{\omega}{{\vk}_0} \right)},
\end{equation}
with $J$ and $Y$ the Bessel functions of the first and second kind, respectively.

Apart from the exact expression of $v_r$ given in  Equation~(\ref{eq:vrgen}), we also can obtain an expression in the WKB approximation, which is instructive to understand the wave properties. Then, by using Equation~(\ref{eq:vrassym}) with the kink velocity given in Equation~(\ref{eq:kinkspeed}), we get velocity
\begin{equation}
 v_r (z) \approx v_r (0)  \exp \left( \frac{z}{4 \Lambda }  \right) \exp \left\{ i 2 \Lambda \frac{\omega}{{\vk}_0} \left[ 1 - \exp \left( - \frac{z}{2 \Lambda} \right) \right] \right\}, \label{eq:vrassymapp}
\end{equation}
and according to Equation~(\ref{eq:wavelength}) the wavelength is
\begin{equation}
\lambda (z) = \frac{2\pi}{\omega} {\vk}_0 \exp \left( \frac{z}{2 \Lambda} \right). \label{eq:wavelengthapp}
\end{equation}
We note that Equation~(\ref{eq:vrassymapp}) is consistent with the condition of validity of the WKB approximation, since it coincides with the expression obtained with the asymptotic expansion of $H_0^{\rm (1)}$ for large arguments, i.e., large $\Lambda$, is used in Equation~(\ref{eq:vrgen}) \citep[see, e.g.,][]{abra}. We see that density stratification increases both amplitude and wavelength of the kink wave as it propagates along the loop from the footpoint to the apex. In the limit of no longitudinal stratification, i.e., $\Lambda \to \infty$, Equations~(\ref{eq:vrassymapp}) and (\ref{eq:wavelengthapp}) simply reduce to
\begin{eqnarray}
 v_r (z) &\approx& v_r (0)  \exp \left( i \frac{\omega}{{\vk}_0} z \right), \\
\lambda (z) &=& \lambda =  \frac{2\pi}{\omega} {\vk}_0  .
\end{eqnarray}
which corresponds to the propagating kink wave in a longitudinally homogeneous tube with constant kink velocity ${\vk}_0$. Both the amplitude and the wavelength are constants in this case.

\begin{figure}[!htp]
\centering
\epsscale{0.99}
\plotone{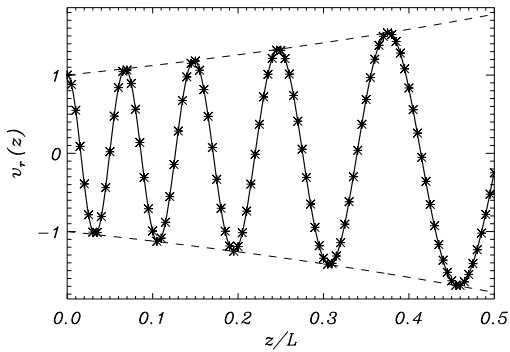}
\caption{Kink wave radial velocity perturbation (solid line), as a function of $z/L$ in a longitudinally stratified tube.  In this case, the transverse transitional layer is absent so there is no resonant damping. The symbols correspond to the WKB approximation (Equation~(\ref{eq:vrassym})). The dashed lines outline the envelope $\exp \left( z / 4 \Lambda \right)$. Results for $\chi = 3$, $\rho_{\rm i 0} /  \rho_{\rm apex} = 10$, and $\omega L / {\vk}_0 = 100 $. \label{fig:comp}}
\end{figure}

We compare in Figure~\ref{fig:comp} the exact $v_r (z)$ given by Equation~(\ref{eq:vrgen}) with the expression in the WKB approximation (Equation~(\ref{eq:vrassymapp})) for a particular set of parameters. We see that there is a very good agreement between both Equations. The effect of longitudinal stratification on both the wavelength and the amplitude is clearly seen in Figure~\ref{fig:comp}.

\subsubsection{Case with Resonant Damping}
\label{exploit}

\begin{figure}[!htp]
\centering
\epsscale{0.99}
\plotone{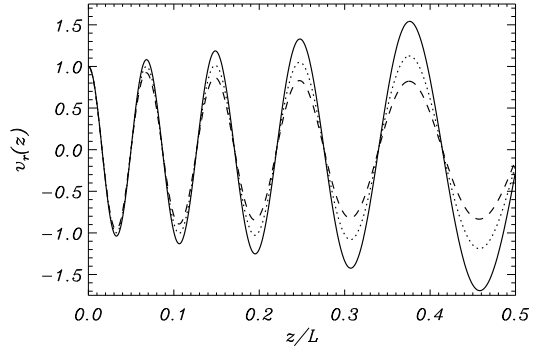}
\caption{Kink wave radial velocity perturbation (solid line), as a function of $z/L$ in a longitudinally stratified tube. The different lines correspond to $l/R = 0$ (solid), $l/R = 0.1$ (dotted), and $l/R = 0.2$ (dashed). In all cases, $\chi = 3$, $\rho_{\rm i 0} /  \rho_{\rm apex} = 10$, and $\omega L / {\vk}_0 = 100 $. \label{fig:reso}}
\end{figure}

Here, we include the effect of damping by resonant absorption. As an exact solution of Equation~(\ref{eq:vrfull}) is very difficult to obtain, we use the expressions derived in the WKB approximation. First of all, the expression for $\lambda (z)$ is the same as given in Equation~(\ref{eq:wavelengthapp}) in the case without resonant damping. Now we compute the damping length, $\ld (z)$, due to resonant absorption using Equation~(\ref{eq:damplen}) with the kink velocity given in Equation~(\ref{eq:kinkspeed}). We obtain
\begin{equation}
 \ld (z) =  2 \pi \frac{\mathcal{F}}{m} \frac{R}{l} \frac{\chi + 1}{ \chi - 1 }\frac{1}{\omega} v_{\rm k 0} \exp \left(  \frac{z}{2 \Lambda} \right). \label{eq:damplenexp}
\end{equation}
We note again that the ratio $\ld (z) / \lambda (z)$ is independent of $z$.

Finally, using Equations~(\ref{eq:vrexp}) and (\ref{eq:amptot}) we obtain the expression for $v_r (z)$ and $\mathcal{A} (z)$, namely

\begin{equation}
 v_r (z) \approx \mathcal{A} (z) \exp \left\{ i 2 \Lambda \frac{\omega}{{\vk}_0}
\left[ 1 - \exp \left( - \frac{z}{2 \Lambda} \right) \right] \right\},
\label{eq:vrexpres}
\end{equation}
with $\mathcal{A} (z)$ the amplitude as a function of $z$ given by
\begin{eqnarray}
 \mathcal{A} (z) &=& v_r\left( 0 \right)  \exp \left( \frac{z}{4 \Lambda} \right) \nonumber \\
&\times& \exp \left\{ - \frac{1}{\pi} \frac{m}{\mathcal{F}} \frac{l}{R} \frac{\chi-1}{\chi+1} \Lambda \frac{\omega}{{\vk}_0} \left[ 1 -  \exp \left( - \frac{z}{2 \Lambda} \right)   \right] \right\}. \label{eq:amptotres}
\end{eqnarray}
In Equation~(\ref{eq:amptotres}) we see that there are two competing effects that determine the amplitude of the propagating kink wave. On the one hand, density stratification causes the amplitude to increase with height; on the other hand, resonant absorption damps the kink wave, so its effect is to decrease the amplitude. Whereas the increase of the amplitude due to longitudinal stratification is independent of $\omega$, the damping by resonant absorption does depend on $\omega$. Thus, we can anticipate that for a particular, critical frequency, $\omega_{\rm crit}$, the amplitude may be a constant independent of $z$. However, due to the functional dependence of Equation~(\ref{eq:amptotres}) on $z$, we clearly see that the amplitude cannot remain constant for all $z$.  Despite  this fact, it is still possible to give an expression for the critical frequency for which the amplitude is a approximately constant for small $z$ only. To do so, we approximate $ 1 -  \exp \left[ - z/(2 \Lambda) \right] \approx  z/(2 \Lambda)$ in Equation~(\ref{eq:amptotres}). Then, we obtain
\begin{equation}
 \omega_{\rm crit} \approx \frac{\pi}{2} \frac{\mathcal{F}}{m} \frac{R}{l} \frac{\chi+1}{\chi-1} \frac{{\vk}_0}{\Lambda}. \label{eq:relationamp0}
\end{equation}
For $\omega \lesssim \omega_{\rm crit}$ longitudinal stratification is the dominant
effect and the amplitude increase with $z$. On the contrary, for $\omega \gtrsim
\omega_{\rm crit}$ damping by resonant absorption is more efficient and the amplitude
decreases in $z$. We must bear in mind that Equation~(\ref{eq:relationamp0}) is only
valid for small $z$, i.e., close to the footpoint of the loop, but gives us some
qualitative information about the behavior of the amplitude depending on the value
of the frequency.

\begin{figure*}[!htp]
\centering
\epsscale{0.49}
\plotone{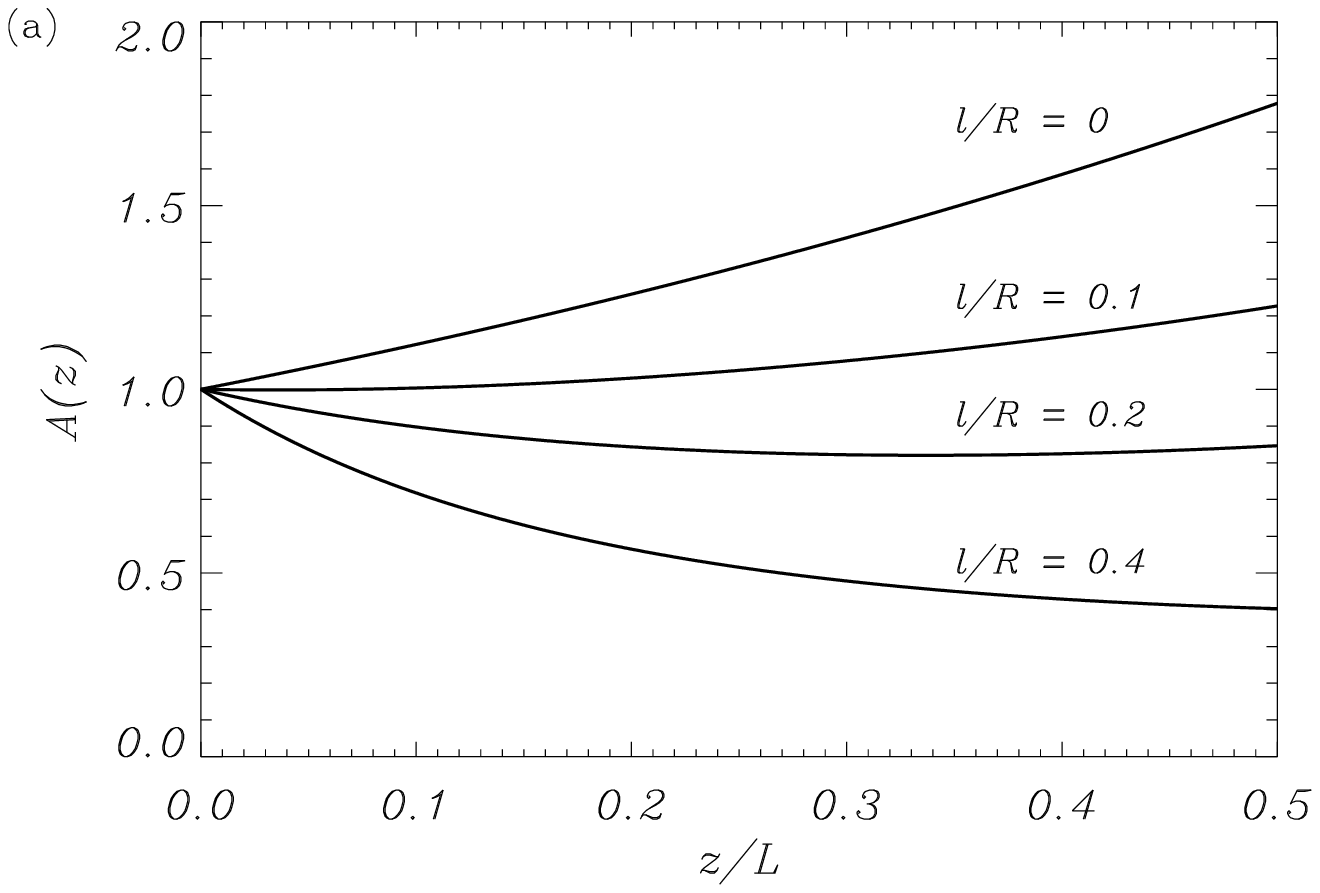}
\plotone{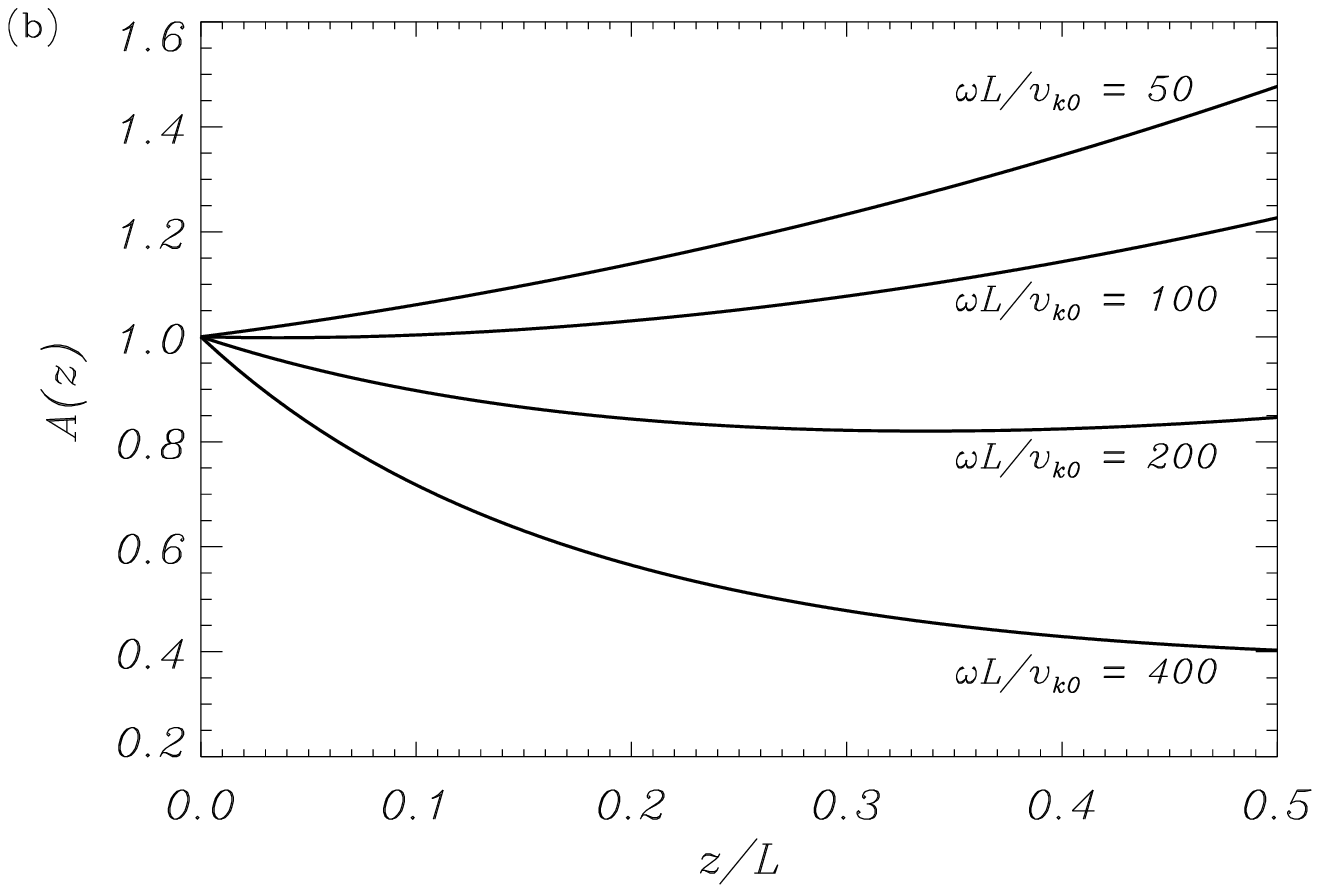}
\plotone{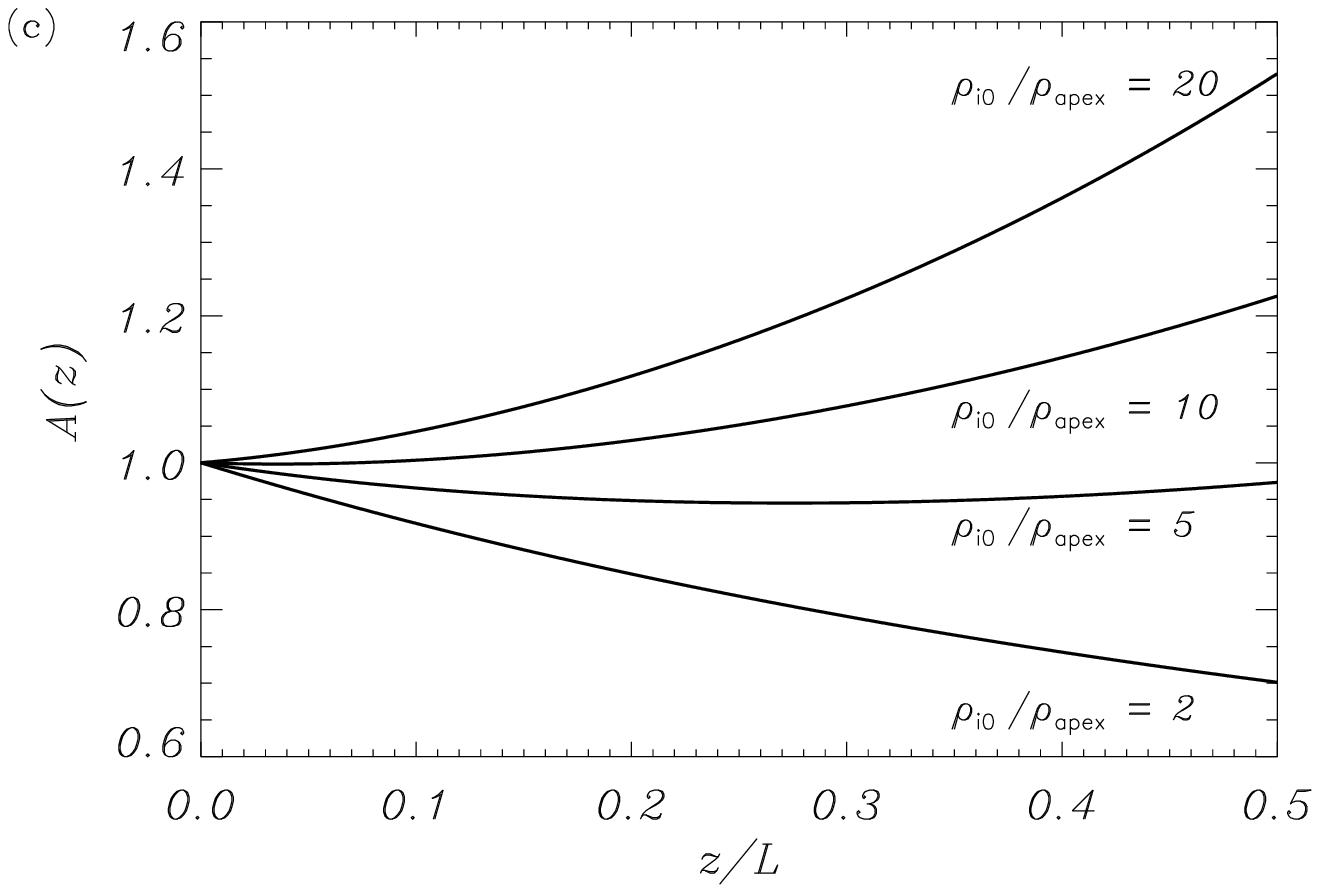}

\caption{Kink wave amplitude as a function of $z/L$ in a longitudinally stratified tube
with a transverse transitional layer. (a) Results for $l/R = 0$, 0.1, 0.2, and, 0.4 with
$\omega L / {\vk}_0 = 100$ and $\rho_{\rm i 0} /  \rho_{\rm apex} = 10$. (b) Results for
$\omega L / {\vk}_0 = 50$, 100, 200, and 400 with $l/R = 0.1$ and $\rho_{\rm i 0} /
\rho_{\rm apex} = 10$. (c) Results for $\rho_{\rm i 0} /  \rho_{\rm apex} = 2$, 5, 10,
and 20  with $\omega L / {\vk}_0 = 100$ and $l/R = 0.1$. In all cases $\chi = 3$.
\label{fig:ampli}}

\end{figure*}

In Figure~\ref{fig:reso} we plot $v_r (z)$ for fixed frequency and longitudinal inhomogeneity scale height, but for different values of $l/R$. In the case $l/R = 0$ there is no resonant damping. As $l/R$ increases, the wave amplitude decreases as a result of resonant absorption. However, we see that the wavelength is independent of the value of $l/R$. This means that, in the TTTB approximation, resonant absorption does not affect the value of the wavelength, which is exclusively determined by the longitudinal inhomogeneity scale height. This is an important result from the observational point of view, because the wavelength carries information about longitudinal stratification only (Equation~(\ref{eq:wavelength})), while the amplitude is influenced by both longitudinal and transverse density profiles (Equation~(\ref{eq:amptot})).

Figure~\ref{fig:ampli} displays the kink wave amplitude computed from
Equation~(\ref{eq:amptotres}) for different combinations of parameters. In
Figure~\ref{fig:ampli}(a) we determine the effect of $l/R$ when the remaining
quantities are kept constant, while Figure~\ref{fig:ampli}(b) and (c) show the effect
of $\omega$ and $\rho_{\rm i 0} /  \rho_{\rm apex}$, respectively. Both $l/R$ and
$\omega$ have a similar effect on the amplitude as both of them control the efficiency
of the resonant damping in front of the increase of the amplitude due to
stratification. When both $l/R$ and $\omega$ get larger, the effect of resonant
damping becomes stronger. On the contrary, $\rho_{\rm i 0} /  \rho_{\rm apex}$
controls the role of longitudinal stratification through $\Lambda$
(Equation~(\ref{eq:deflamstrat})). As $\rho_{\rm i 0} /  \rho_{\rm apex}$ grows
($\Lambda$ decreases), the increase in amplitude due to stratification becomes
more important. For the set of parameters used in Figure~\ref{fig:ampli}(b), the
approximate critical frequency for a constant amplitude according to
Equation~(\ref{eq:relationamp0}) is $\omega_{\rm crit} L / {\vk}_0 \approx 100$. We
see that, as expected, for the critical frequency, amplitude is
constant for small $z$, but then increases slightly with larger $z$. We stress again that, for fixed frequency, the amplitude is determined by the combined effect of longitudinal stratification and transverse inhomogeneity, while the wavelength is only affected by longitudinal stratification.

\subsection{Time-dependent Numerical Simulations}

Here the aim is to show how the analytical results derived in the previous
sections for the eigenmode problem are related to the time-dependent problem. In addition, we go beyond the
TT and TB approximations used in the derivation of the
analytical expressions, allowing us further insight. To study the problem of linear propagating waves,
Equations~(\ref{eq:b1}) and~(\ref{eq:b2}) are numerically solved in cylindrical
coordinates. Since we are only interested in the kink mode we assume an azimuthal
dependence of the form $\exp \left( i m \varphi\right)$ with $m=1$, and the equations are
integrated in the radial and  longitudinal coordinates ($r$ and $z$ in our case, respectively). A driver at
the footpoint of the loop ($z=0$), with a single frequency $\omega$, is introduced to
excite propagating waves along the tube. The spatial form of the driver in the
radial direction is chosen, to simplify things, to be the eigenmode of a loop with no resonant layer. This enables us to mainly excite the resonantly damped
modes, while the direct excitation of Alfv\'en modes is very weak. Thus,
the driver at $z=0$ is implemented through the periodic variation in time and the
radial dependence of the kink mode eigenfunctions.  The simulations are done with the code MoLMHD \citep[see][for
further details about the numerical method]{bonaetal10} in a uniform grid of
$1000\times100$ points in the $r$-$z$ plane, and in the range $0<r/R<10$ and
$0<z/R<100$. For convenience, in the code time is expressed in units of $\ta = R / v_{\rm A i 0}$, with $v_{\rm A i 0}$ the internal Alfv\'en velocity at $z=0$, so that the dimensionless frequency is $\omega \ta$. Note that in the previous sections, the dimensionless frequency has been written as $\omega \tau_{\rm k}$, with $\tau_{\rm k} = L / {\vk}_0$. The relation between the two time scales is $\ta = \tau_{\rm k} R/L \sqrt{2\chi/(1+\chi)}$.

\begin{figure*}[!htp] \centering \epsscale{0.49}
\plotone{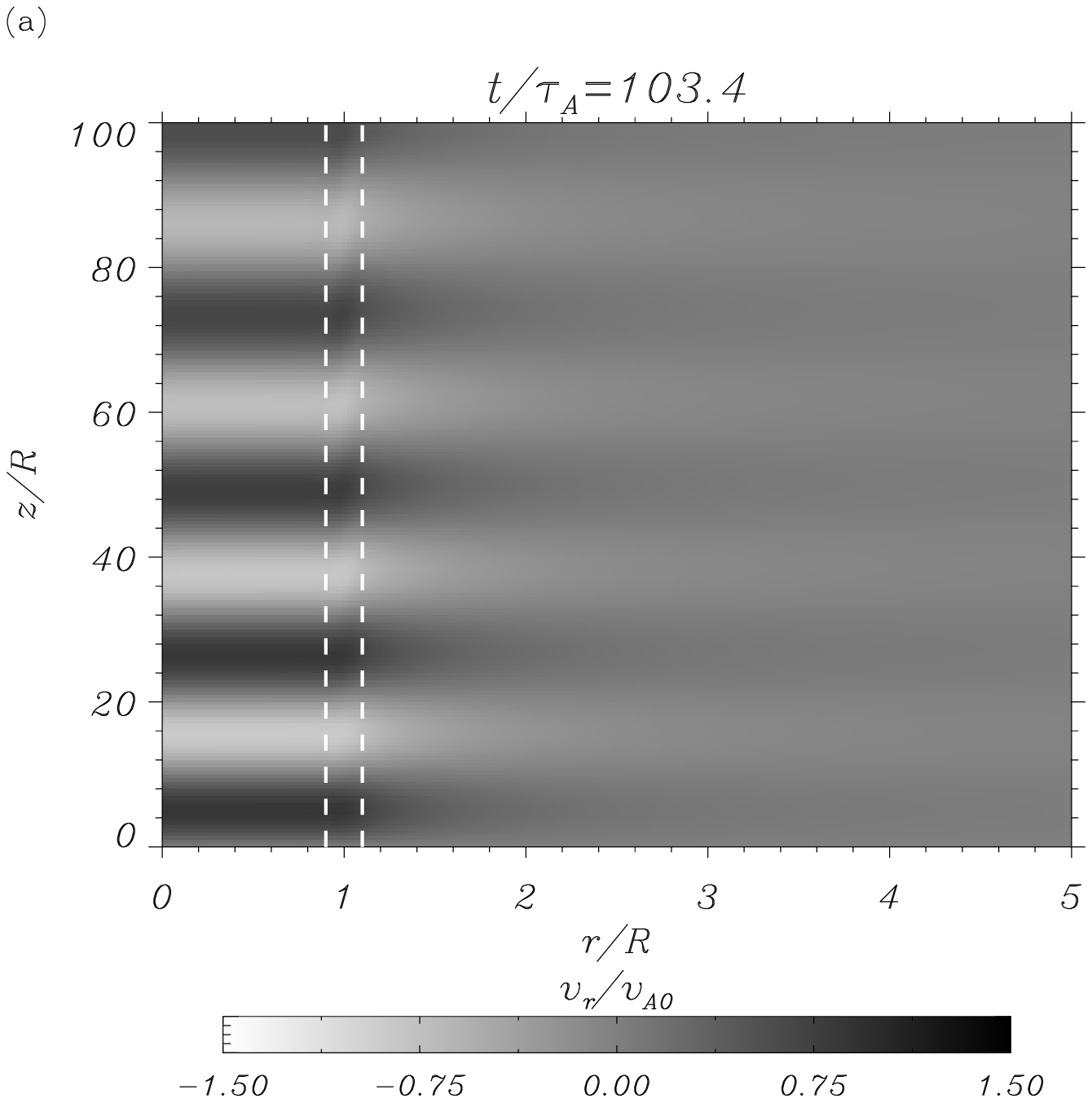}
\plotone{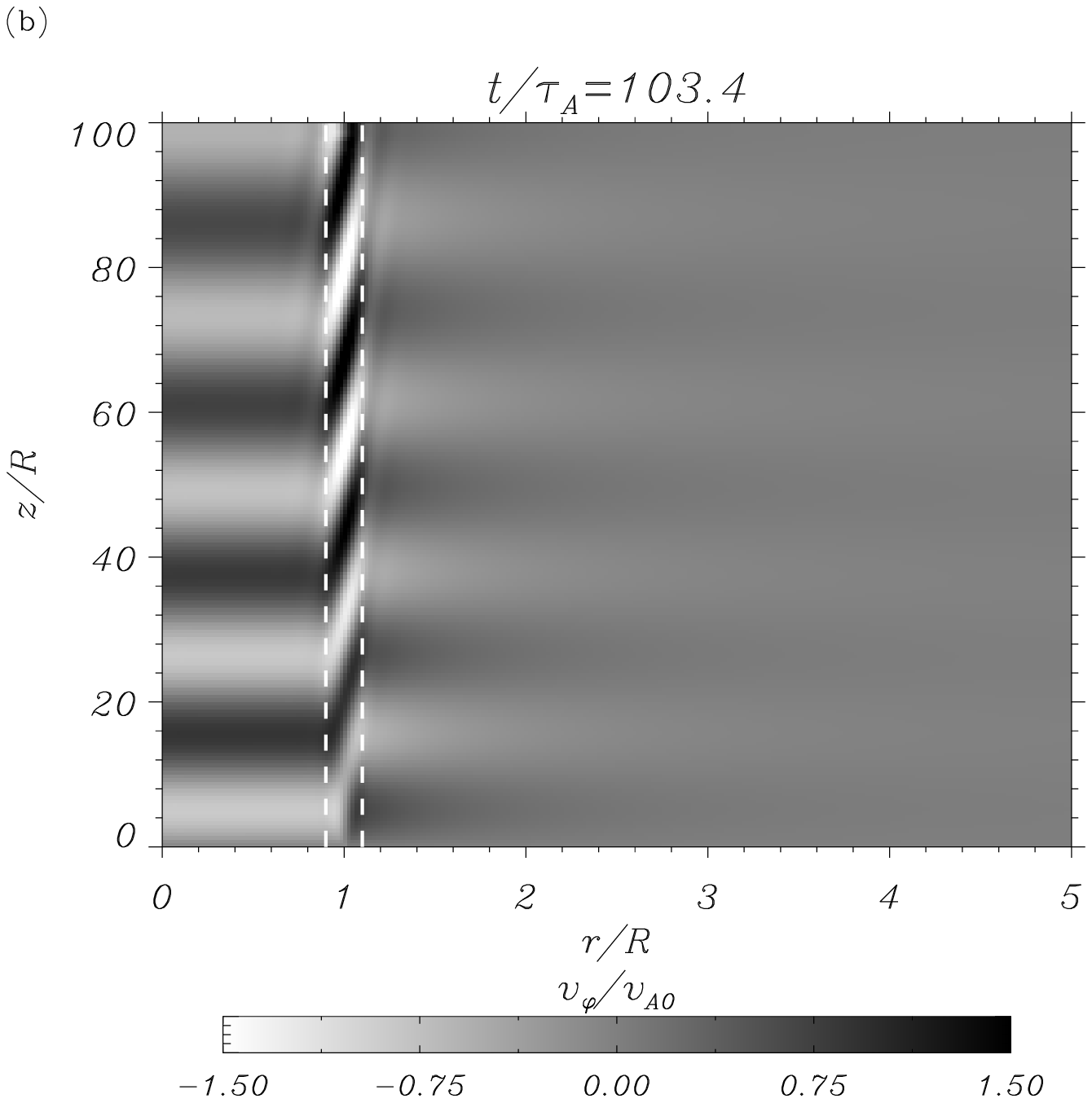}
\caption{
Snapshot of the velocity perturbations (a) $v_r$ and (b) $v_\varphi$ at a given time (indicated on top of the panels).  The driver
is located at $z=0$ and its frequency is $\omega \ta = 0.36$. In this plot the
width of the transverse transitional layer (see dashed lines) is $l/R = 0.2$, $\chi = 3$, and
$\Lambda/R=200$. Note the different spatial scale in the $r$ and $z$ axes. \label{fig:vr_vphi}}
\end{figure*}

\begin{figure}[!htp]\epsscale{0.99}
\plotone{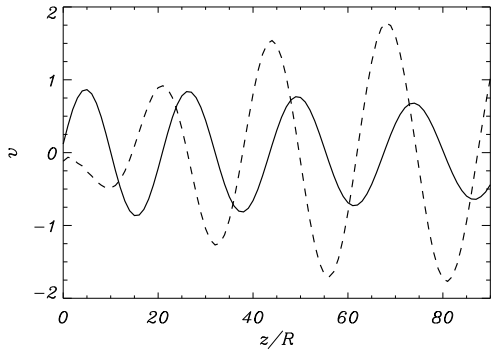}
\caption{
Velocity perturbations $v_r$ (solid line) at $r=0$ and $v_\varphi$ (dashed line) at
$r=R$ as functions of the position along the loop at a given time ($t/\tau_{\rm{A}}=103.4$).
These results correspond to the simulation shown in Figure~\ref{fig:vr_vphi}.
\label{fig:vr_vphi_z}}
\end{figure}

Hereafter, we concentrate on the analysis of the two main variables, $v_r$
and $v_\varphi$, i.e., the radial and azimuthal velocity perturbations, respectively. Figure~\ref{fig:vr_vphi} shows the two-dimensional spatial distribution of the two
velocity components, $v_r$ and  $v_\varphi$, at a given time and for a particular set
of parameters. The driver excites propagating waves that move upwards along
the tube with a characteristic wavelength and amplitude. The
amplitude inside the tube decreases with height for both $v_r$ and $v_\varphi$,
while the $v_\varphi$ component shows a significant increase in $z$ within the
inhomogeneous layer \citep[see the work by][for a longitudinally homogeneous loop]{pascoeetal10}. In  Figure~\ref{fig:vr_vphi_z},
we perform two cuts along the $z$-direction of $v_r$ at $r=0$ and $v_\varphi$ at $r=R$. The radial velocity perturbation at the center of the loop decreases its
amplitude in $z$ due to the combined effect of longitudinal stratification and resonant
absorption. On the contrary, the amplitude of $v_\varphi$  at $r=R$ grows in $z$ because of the conversion from radial to azimuthal motions that takes
place in the inhomogeneous layer due to the Alfv\'en resonance. As a consequence of this
process more energy is concentrated in the layer in the form of azimuthal motions as the wave propagates upwards along the
loop.

\begin{figure}[!htp]\epsscale{0.99}
\plotone{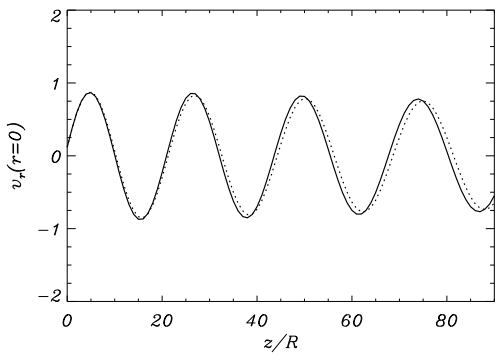}
\caption{
Simulated radial velocity perturbation (solid line) and analytical curve (dotted line) at $r=0$
as functions of position along the loop at $t/\ta=103.4$. In
this plot $l/R = 0.1$, $\Lambda/R=200$ and $\omega \ta = 0.36$.
\label{fig:vr_L200l_0.1}}
\end{figure}

First, we perform some test cases in order to compare the full numerical solutions with the analytical expressions in the TTTB and WKB approximation. We start with an
example that satisfies both the TTTB and WKB approximations. In this example the
frequency of the driver is $\omega \ta = 0.36$. The transitional layer thickness is set
to $l/R=0.1$, i.e., we consider a a thin layer, and the stratification scale height is $\Lambda/R=200$. Waves with a
 wavelength much longer that the tube radius and much shorter than the stratification scale height are excited. Therefore, we are in the TTTB and WKB regimes. In Figure~\ref{fig:vr_L200l_0.1} we plot the $z$-dependence of $v_r$ at $r=0$. The radial velocity perturbation shows an almost constant amplitude in $z$, while the wavelength increases slightly with height. In this particular example, the effect of longitudinal stratification and resonant absorption on the amplitude cancel each other.  The numerical curve is in quite good agreement with the analytical expression given by Equation~(\ref{eq:vrexpres}). Thus, this simulation
shows that the assumptions made to derive Equation~(\ref{eq:vrexpres}) are
well justified when $R \ll \lambda \ll \Lambda$. The almost negligible differences between the numerical and analytical results become smaller as we take smaller values of $R$ and larger values of $\Lambda$.

\begin{figure}[!htp]\epsscale{0.99}
\plotone{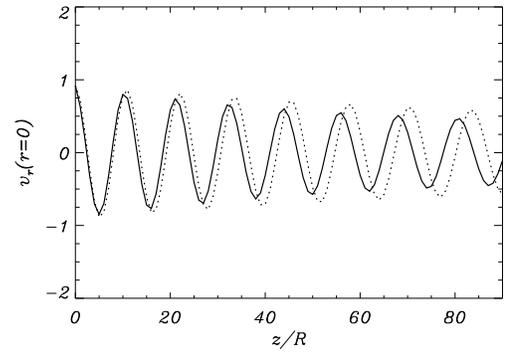}
\caption{
Simulated radial velocity perturbation (solid line) and analytical curve (dotted line) at $r=0$
as functions of position along the loop at $t/\ta =103.4$. In
this plot $l/R = 0.1$, $\Lambda/R=200$ and $\omega \ta = 0.72$.
\label{fig:vr_L200l_0.1w}}
\end{figure}

\begin{figure}[!htp]\epsscale{0.99}
\plotone{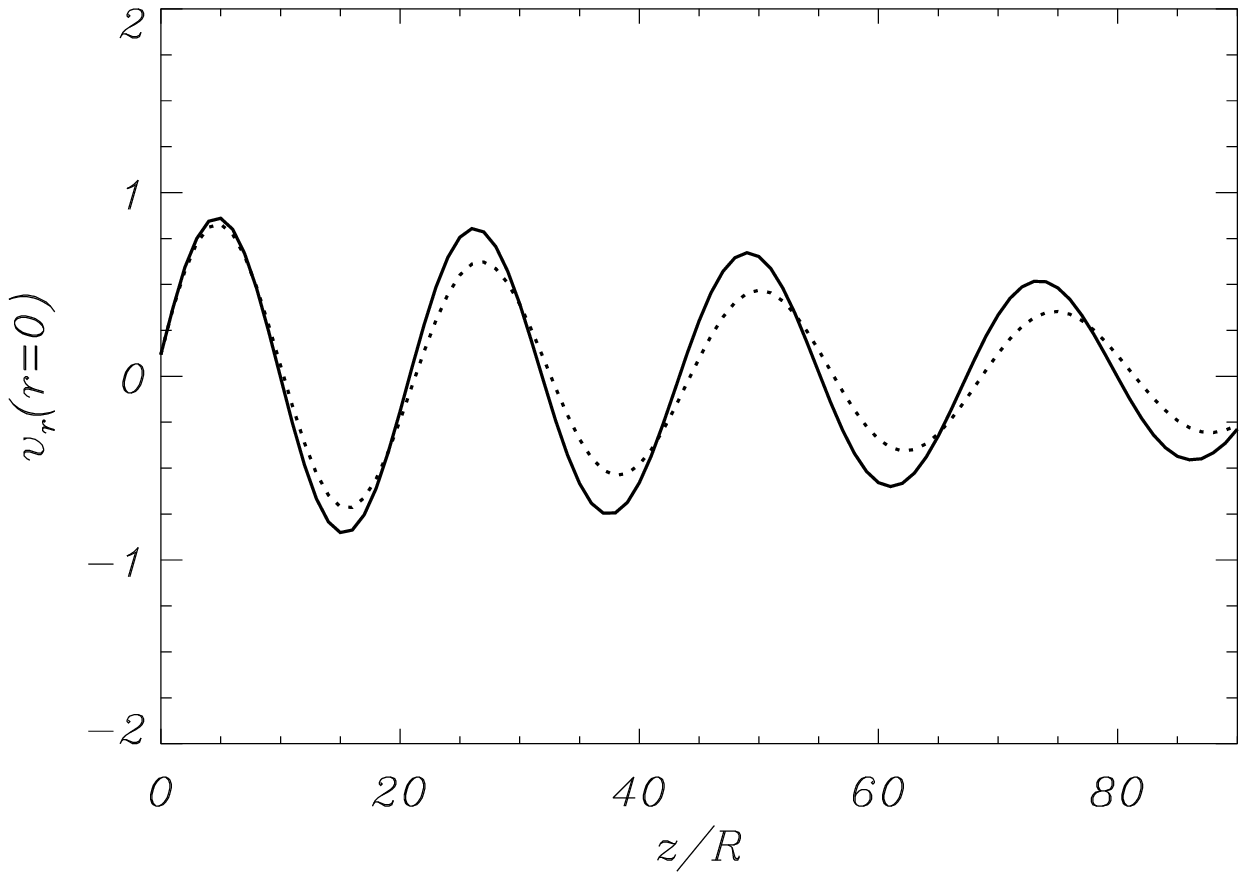}
\caption{
Simulated radial velocity perturbation (solid line) and analytical curve (dotted line) at $r=0$
as functions of position along the loop at $t/\ta=103.4$. In
this plot $l/R = 0.4$, $\Lambda/R=200$, and $\omega \ta = 0.36$.
\label{fig:vr_L200l_0.4}}
\end{figure}

\begin{figure}[!htp]\epsscale{0.99}
\plotone{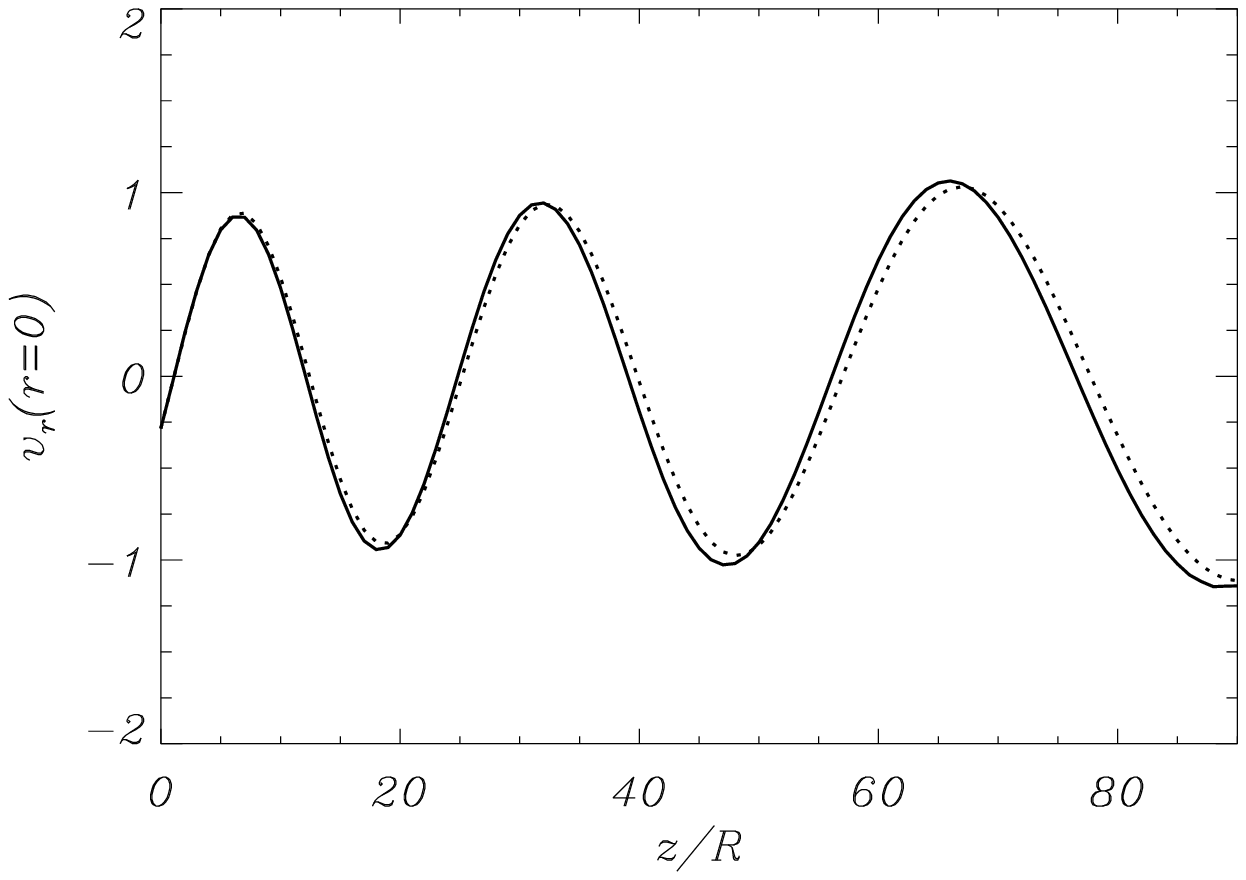} \caption{ Simulated
radial velocity perturbation (solid line) and analytical curve (dotted line) at $r=0$
as functions of position along the loop at $t/\ta =70.1$.
In this plot $l/R = 0.1$, $\Lambda/R=50$, and $\omega \ta = 0.36$.
\label{fig:vr_L50l_0.1}} \end{figure}

\begin{figure}[!htp]\epsscale{0.99}
\plotone{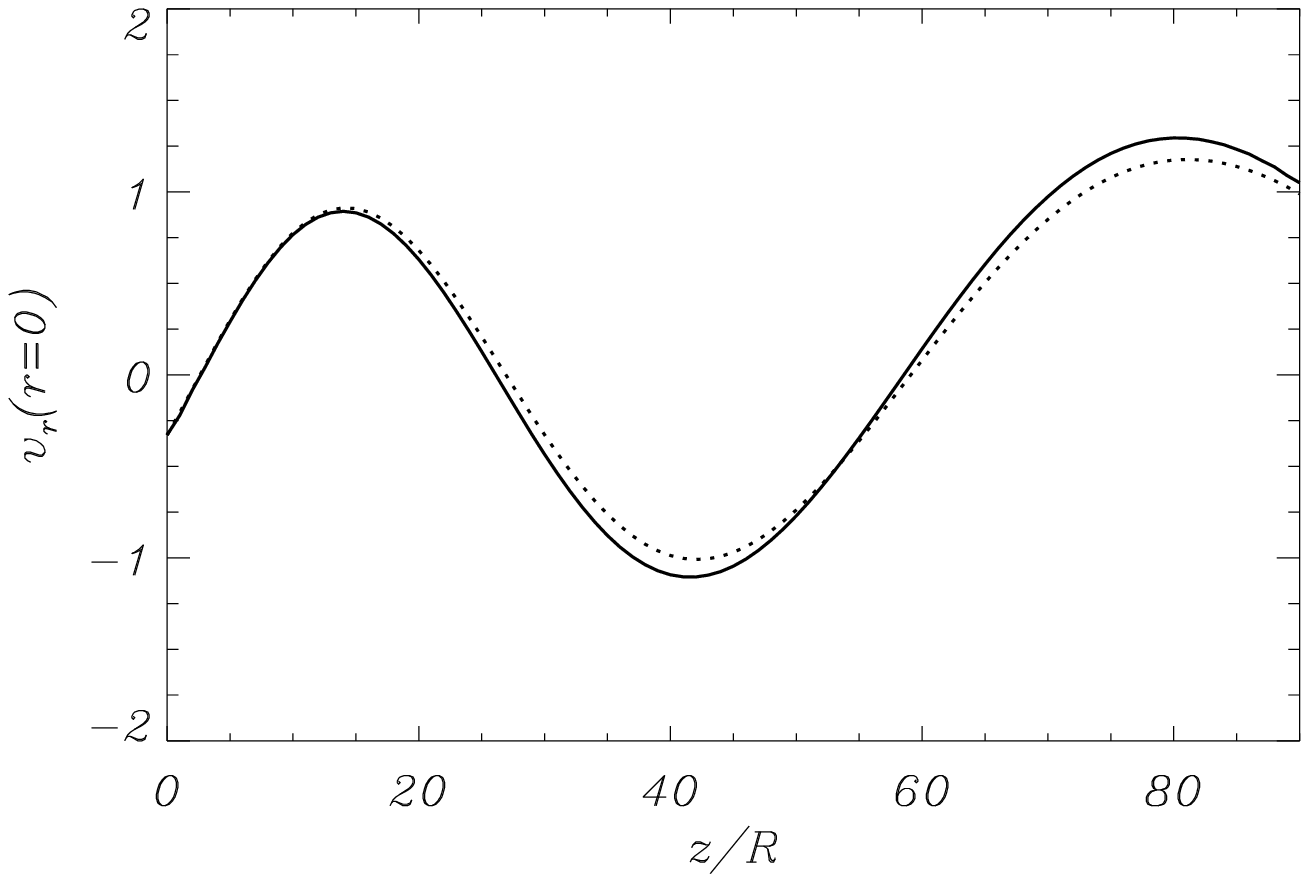} \caption{ Same as Figure~\ref{fig:vr_L50l_0.1} but for $\omega \ta = 0.18$.
\label{fig:10}} \end{figure}

Next, we show how the results are modified when we depart from the valid regimes of the TTTB and WKB approximations. In a second example (see Figure~\ref{fig:vr_L200l_0.1w}) the driver frequency is $\omega \ta = 0.72$ and the rest of parameters are the same as in the previous case, but now the excited kink wavelength is almost twice smaller than before and the assumption of TT may be compromised. The comparison of the numerical result and analytical TTTB expression indicates that for large $z$  the analytical formula overestimates both amplitude of oscillation and wavelength. In our third example, we return to the case with $\omega \ta = 0.36$ but we use a larger value of the transitional layer thickness, $l/R=0.4$. Then, we are not in the TB regime. The results in this case (see Figure~\ref{fig:vr_L200l_0.4}) point out that the analytical expressions predict a stronger damping of amplitude and a larger increase of wavelength with $z$ in comparison with our numerical results. In our last case, we test the WKB approximation by considering that the wavelength of the excited kink mode is not much shorter than  the stratification scale height. First we take in Figure~\ref{fig:vr_L50l_0.1} the same parameters as in Figure~\ref{fig:vr_L200l_0.1} but with $\Lambda/R=50$ and $\omega \ta = 0.36$. Now the wavelength and the stratification scale height are of the same order, approximately, and we are not in the WKB regime strictly. Next in Figure~\ref{fig:10} we use exactly the same parameters but $\omega \ta = 0.18$, so that the wavelength is longer than the stratification scale height and the condition of applicability of the WKB approximation is violated. However, both Figures~\ref{fig:vr_L50l_0.1} and \ref{fig:10} show that the full numerical solutions and the WKB results are in remarkable good agreement.

The  numerical simulations performed in this Section tell us that the analytical expressions derived in the TTTB and WKB approximations are very accurate when kink wavelength is in the range  $R \ll \lambda \ll \Lambda$ and $l/R \ll 1$. We also obtain a very good agreement when the condition $\lambda \ll \Lambda$ is relaxed. For thick layers, i.e., outside the TB approximation, the analytical expression for the amplitude gives a slightly stronger damping than in the full numerical case, although the wavelength remains approximately correct. If instead we depart from the TT approximation, the analytical approximation of amplitude and wavelength gives slightly larger values in comparison with the numerical results. However, as can be seen in Figures~\ref{fig:vr_L200l_0.1w}--\ref{fig:10},  the differences between the numerical simulations and the analytical predictions are remarkably small even in the worst case scenarios studied here.

\section{IMPLICATIONS FOR SOLAR ATMOSPHERIC MAGNETOSEISMOLOGY}

\label{sec:seis}

The theoretical results of this paper show that determining the longitudinal and transverse inhomogeneity length scales of solar waveguides from observations of propagating kink waves is not a trivial matter. As was shown in Section \ref{exploit}, if we can observationally identify the wavelength (or phase speed) as a function of $z$, we know this is independent of the transverse inhomogeneity. Therefore, if we can simultaneously estimate the wavelength (or phase speed) and amplitude of the kink wave as a function of $z$ we will have enough information to quantify the possible contribution of resonant absorption on attenuating the observed amplitude. However, observationally this presents difficult challenges. In the WKB approximation, for a flux tube with density stratification, constant magnetic field and no resonant layer, Equation (\ref{eq:vrassym}) shows that the velocity amplitude is related to the phase speed by
\begin{equation}
v_r (z) \propto \sqrt{\vk (z)}.
\label{undamped}
\end{equation}
Hence if this relationship does not hold for an observed wave, the discrepancy could be caused by damping due to the presence of a resonant layer, i.e., transverse inhomogeneity. At present a possible avenue to for this type of magnetoseismology is by using the high spatial/temporal resolution Solar Optical Telescope (SOT) onboard Hinode. Using the Ca II broadband filter of SOT chromospheric kink waves propagating in spicules can be analyzed. As was shown by \citet{he2009b}, it is possible to track both the propagating kink wave amplitudes and phase speeds from the photosphere up to the lower corona. \citet{he2009a} have also shown time snapshots of these propagating waves that nicely show the relationship between wavelength and amplitude with height. Unfortunately, since there are only a few wavelengths detected up to the visible apex of spicules it is difficult to detect a clear trend in the variation of wavelength with height. Therefore a more practical approach of estimating $\vk(z)$ is by tracking the phase travel time of the kink wave as was done by \citet{he2009b}.

Regarding CoMP data, \citet{VTG} attempted to estimate the damping length of kink waves propagating along coronal loops, assuming the plasma was longitudinally homogeneous. To interpret CoMP data, assuming there is longitudinal variation in plasma density, as stated previously, we need both information about velocity amplitude and phase speed to see if Equation (\ref{undamped}) holds. Certainly, \citet{tomczyk09} showed that it was possible to estimate $\vk(z)$ from their time-distance plots \citep[see Figure 5 in][]{tomczyk09}. However, since the velocity amplitudes were small (only a few km $^{-1}$s) it will be more of a challenge to estimate the longitudinal variation of this quantity.

We must also mention the physical limitation of using a model flux tube with constant magnetic field. It was shown by \citet{ruderman2008} that if the magnetic field is varying longitudinally (still with density stratification), in the thin tube approximation Equation (\ref{undamped}) becomes
\begin{equation}
v_r (z) \propto R(z) \sqrt{\vk (z)},
\label{Rundamped}
\end{equation}
where $R(z)$ is the flux tube radius. Indeed, Equation (\ref{Rundamped}) was used by \citet{verth2011} as the basis for interpreting propagating kink waves in a spicules to determine chromospheric flux tube expansion, since both $v_r(z)$ and $\vk(z)$ were estimated from observation. It can be seen from Equation (\ref{Rundamped}) that if there is a resonant layer present, assuming ideal MHD will cause the expansion of flux tube with height to be underestimated. Of course, Equation (\ref{Rundamped}) can also be applied to CoMP observations of propagating kink waves to estimate the variation of the magnetic field along coronal loops since in the thin tube flux conservation is given by $B(z) \propto 1/R^2(z)$. Magnetoseismology of post-flare standing kink waves in coronal loops, allowing for both longitudinal variation in magnetic field strength and plasma density was first done by \citet{verth2008}, therefore it would be interesting to see if propagating coronal kink waves could be exploited for a similar purpose.

\section{DISCUSSION AND CONCLUSIONS}

\label{sec:discussion}

In this paper we have investigated the propagation of resonantly damped kink MHD waves in
both transversely and longitudinally inhomogeneous solar waveguides. The present
work extends the previous investigation by TGV, where longitudinal inhomogeneity was
not taken into account. By using the WKB method, we have derived general expressions
for amplitude and wavelength of kink waves in the TT and TB
approximations. Variation of wavelength along the magnetic flux tube only
depends on longitudinal stratification. However, wave amplitude is affected by
both transverse and longitudinal inhomogeneity. The kink mode is
damped by resonant absorption in the Alfv\'en continuum due to transverse
inhomogeneity and the damping length changes along the waveguide since it is dependent on the longitudinal density profile. However, it is important to note that the ratio of damping length to wavelength
along the magnetic flux tube is constant. It is obvious that the effect of longitudinal
inhomogeneity can both amplify or attenuate the wave, depending on whether the density
decreases or increases towards the direction of wave propagation, respectively. In the case of radial and longitudinal inhomogeneity we have shown that the
variation of kink mode amplitude along magnetic tubes is determined
by the combined effect of both these mechanisms.

We have performed an application of the theory to the case of driven kink waves in stratified coronal loops
propagating upward from the loop footpoints. In this model the density decreases
exponentially along the flux tube, and so resonant absorption and longitudinal
stratification have opposite effects on the kink wave amplitude. The efficiency of
resonant absorption as a damping mechanism depends on the frequency. For frequencies
larger than an approximate critical value, the net effect is the decrease of the
amplitude in $z$. The opposite effect, i.e., increase of the amplitude, takes place for
frequencies smaller than the approximate critical one. These analytical results have been
checked using full numerical simulations beyond the TTTB and WKB approximations. The analytical
expressions and the full numerical solution of driven waves are in good agreement even when the requirements of the TTTB and WKB approximations are not strictly fulfilled.

The analytical expressions for the wavelength and the amplitude derived in this paper
have direct implications for solar magnetoseismology of, e.g., coronal loops, spicules, and prominence threads, allowing the equilibrium model to have realistic variation in plasma density in both the longitudinal and transverse direction to the magnetic field. This is an important step forward for exploiting the observations of ubiquitous propagating kink waves to probe the plasma fine structure of solar atmosphere by implementing magnetoseismological techniques.

\acknowledgements{RS acknowledges support from a postdoctoral fellowship within the EU
Research and Training Network ``SOLAIRE'' (MTRN-CT-2006-035484). JT  acknowledges support
from the Spanish Ministerio de Educaci\'on y Ciencia through a Ram\'on y Cajal grant and
funding provided under projects AYA2006-07637 and FEDER funds. GV and MG acknowledge
support from K.U. Leuven via GOA/2009-009.}

\end{document}